# Hydrogen Surfactant Assisted Coherent Growth of GaN on ZnO Substrate

*Jingzhao Zhang, Yiou Zhang, Kinfai Tse and Junyi Zhu[1]*

*Department of Physics, the Chinese University of Hong Kong, Shatin, New Territories, Hong Kong, China*

**Abstract**

Heterostructures of wurtzite based devices have attracted great research interests since the tremendous success of GaN in light emitting diodes (LED) industry. Among the possible heterostructure material candidates, high quality GaN thin films on inexpensive and lattice matched ZnO substrate are both commercially and technologically desirable. However, the energy of ZnO polar surfaces is much lower than that of GaN polar surfaces. Therefore, the intrinsic wetting condition forbids such heterostructures. As a result, poor crystal quality and 3D growth mode were obtained. To dramatically change the growth mode of the heterostructure, we propose to use hydrogen as a surfactant, confirmed by our first principles calculations. Stable H involved surface configurations and interfaces are investigated, with the help of newly developed algorithms. By applying the experimental Gibbs free energy of $H_2$, we also predict the temperature and chemical potential of H, which is critical in experimental realizations of our strategy. This novel approach will for the first time make the growth of high quality GaN thin films on ZnO substrates possible. We believe that our new strategy may reduce the manufactory cost and improve the crystal quality and the efficiency of GaN based devices.

## I. Introduction

Semiconductors with wurtzite (WZ) structures have broad applications in modern semiconductor industry [1-6]. ZnO and GaN are two representative compounds. However, the practical device application using ZnO is still undergoing inherent problems, e.g. the major difficulty in the *p*-type doping. Meanwhile, lacking of low cost and lattice matched substrates remains a big challenge for many group III-nitrides [7-9].

Devices based on ZnO/GaN heterostructures [3,10-12] have drawn research interests because of their similarities in the crystal structures, growth direction, and lattice constants. The lattice mismatch ratio is less than 2% [13,14] (ZnO is slightly larger). In recent years, ZnO wafers made from high-quality and large single crystalline have been reported, using hydrothermal growth method [15-17]. This method is currently thought to be at the threshold of real mass-production of ZnO crystals [18,19], indicating the availability of commercialization [20]. Additionally, ZnO substrate may decrease the usage of expensive gallium for buffer layers. Therefore, ZnO has been proposed to be an ideal substrate [21] or buffer layers [22] to grow GaN.

Many experimental attempts have been taken to grow high crystal quality GaN on ZnO substrates [12,23-29]. However, experimentally, it is difficult to form layer by layer type coherent growth, especially for molecular beam epitaxy and organometallic vapor phase epitaxy techniques at high temperature. Instead, 3D-like growth have been observed [12,28], with some interface layers of secondary phases like $Ga_2O_3$, $Ga_2ZnO_4$ and $Zn_3N_2$ [12,30-32]. Moreover, it has been shown that even if the secondary phase layers are avoided, the growth was still 3D-like [12]. On the contrary, it is generally easier to grow ZnO on GaN substrates [30,31,33-35]. On top of a single crystal GaN substrate, ZnO crystal films of high quality and sharp interfaces have been observed [31,34].

---

[1] Email: jyzhu@phy.cuhk.edu.hk

Theoretical analysis on these experimental observations is essential yet lacked. Our previous theoretical works focused on the unreconstructed surface energy calculations and no studies on the interfaces have been included [36]. The intrinsic difficulty is the estimation of the interface energy. As we shall show in the latter part of this paper, a new algorithm to estimate the interface energy is proposed with high accuracies. In Ref. [36], it shows that the surface energies of GaN polar surfaces are generally much larger than that of ZnO. The differences are up to several tens of meV/Å$^2$, indicating the difficulties to make GaN wet ZnO.

To modify the unfavorable growth mode, surfactants can be effective [37,38]. Surfactant studies have been mostly focused on large metallic elements [37-43]. For example, arsenic as a surfactant was introduced in Si/Ge/Si(001) heterostructures to change growth modes in the late 1980s [37,38]. This is because large, nonvolatile, metallic elements are likely to float on top of the growth front.

Whether small elements like hydrogen can serve as surfactants is an interesting and open question in the growth of GaN or ZnO [39,40,44-46]. Limited studies of H as surfactants were reported in a few other semiconductors. H as a surfactant may enhance the interchange mechanism in Si/Ge heterostructures [47]. H and Sb as dual surfactants were reported to enhance Zn incorporation in GaP [39]. During the standard OMVPE growth processes, hydrogen is a common unavoidable impurity [48,49], often decomposed from precursors or carrier gases [50]. As a result, hydrogen-involved surface reconstructions have also been reported on the polar surfaces of ZnO [51-55] and GaN [56,57] both in experiments and theories. Such surface reconstructions are usually stable because H may help the surface to satisfy electron-counting-rule (ECR) [58]. However, when H atoms are adsorbed on the surface, they may either desorb or incorporate into the bulk. Therefore, to use H as a surfactant during epitaxial growth, a hydrogen source with proper partial pressure will be crucial in the growth environment.

Despite these disadvantages, hydrogen has advantages over large metallic surfactants: (1) it is more flexible to passivate different surfaces of various materials and satisfy ECR; (2) those hydrogen atoms diffused into bulk may passivate dangling bonds in the bulk or on the interfaces and improve the crystal quality by suppressing the formation of certain deep defects [49,59,60]; (3) H can be easily driven out by post-annealing treatment [61].

Nevertheless, it is a great technological challenge to control the growth condition and incorporate proper amount of H during the growth. It's essential to perform a calculation of the vibrational entropy of H, which is the major component of the chemical potential of H during high temperature growth [52-56,62-66]. In addition to such theoretical calculation, applying the experimental Gibbs free energy of $H_2$ is also a possible alternative method to obtain the quantitative relations of H chemical potential, temperature ($T$) and pressure ($p$). However, such practices are rare in the theoretical analysis except a few works [40]. In addition to the temperature dependent term of H chemical potential, hydrogen passivation on different surfaces of ZnO and GaN involves different numbers of H atoms, which may complicate the whole analysis. Therefore, the wetting condition can be very sensitive to the H incorporation and its atom numbers. In this paper, we obtained the correct range of H chemical potential under typical growth temperature and pressure to satisfy the desired layer by layer growth condition. In addition to the thermodynamic analysis, a kinetic study of H on the surface or in the bulk can be also very important. Yet, it's out of the scope of this paper.

Also, due to the sensitive nature of the H chemical potential dependence on the growth mode, an accurate estimation of the absolute surface energy of the substrate materials, that of the epi-layers, and the absolute interface energy is needed [37]. In this paper, we followed our previously developed methods to estimate the absolute surface energies and also devised new strategies to calculate the absolute interface energy [67,68]. The detailed discussions are shown in the methodology part. Also, this is the first time that such complex analysis has ever been applied to a heterostructure along polar surfaces. Meanwhile, we calculated the phase diagram of a sharp ZnO/GaN interface. The wetting condition as a function of H chemical potential is obtained. Finally, based on all the above analysis, we obtained accurate predictions of proper growth conditions so that coherent growth of GaN thin film on ZnO substrate can be achieved.

## II. Methodology

The growth mode is determined by the free energy of the substrate surface ($\sigma_s$), the interface free energy ($\sigma_i$), and the surface free energy of the hetero-epitaxial layer ($\sigma_f$). The inequality:

$$\sigma_s > \sigma_i + \sigma_f, \tag{1}$$

sets the condition for the epitaxial film to wet the substrate [37]. Therefore, the proper wetting condition can be achieved if we can calculate all these three energies accurately.

To estimate the absolute surface energies of GaN and ZnO, we adopted our recent approach that yields relatively accurate absolute surface energies of zincblende $(111)/(\bar{1}\bar{1}\bar{1})$ and Wurtzite $(0001)/(000\bar{1})$ surfaces [36,37]. The estimated errors are in the order of a few meV/Å$^2$ [36]. Still, the early calculations were performed mainly on non-reconstructed surfaces. The accurate prediction of growth mode will need the absolute energy based on surface reconstructions. In this paper, we adopted surface reconstructions proposed by early STM observations and first principles calculations[53,57,69,70]. In this paper, we follow our cluster and pseudo molecule method [36] strictly.

The remaining challenge will be how to calculate the accurate interface energy. Here we constructed a simple slab model to calculate the interface energies of various interface configurations, especially for WZ materials. The structure is shown in **Fig. 1**. Both of the top and bottom surfaces are passivated with fractional-charged pseudo-H [71]. Since we can obtain the pseudo chemical potential (PCPs) of the pseudo hydrogen accurately, which are defined in our previous works [36,72], it's straightforward to further obtain the absolute interface energy by deducting the energy contributions from both surfaces. For any arbitrary interface configurations, absolute interface energies can be determined by the following equation:

$$\sigma_i = \frac{1}{\alpha_{\text{interface}}}(E_{\text{interface}} - \sum_i n_i \cdot \mu_i - n_{H_A}\hat{\mu}_{H_A} - n_{H_B}\hat{\mu}_{H_B}) \qquad (2)$$

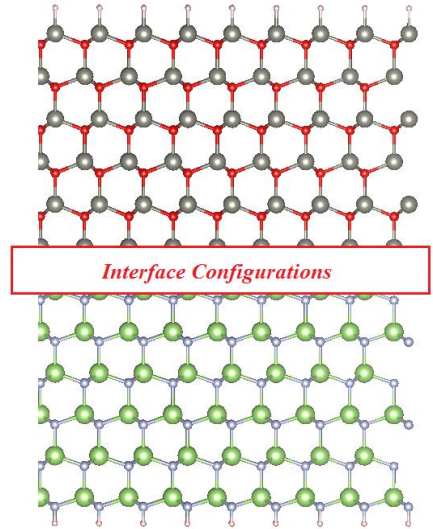

**FIG. 1**. A schematic illustration to the strategy of interface energy calculations, the bottoms on both sides are fully passivated with pseudo-H.

where $\alpha_{\text{interface}}$ is the area of the interface, $E_{\text{interface}}$ is the total energy of the fully passivated slab, shown in **Fig. 1**, $n_i$ and $\mu_i$ are the numbers and chemical potentials of the slab atoms (e.g. Zn, O, Ga, N), $n_{H_A}$ and $n_{H_B}$ are the number of pseudo-H atoms used to passivate the bottom and top surfaces of the slab, and $\hat{\mu}_{H_A}$, $\hat{\mu}_{H_B}$ are the PCPs of pseudo-H. Other interfaces with different polarities and configurations can be constructed in similar ways, and the obtained absolute interface energies are directly comparable. This universal approach is quite different from the former modeling on the interface energy calculations, which focused on superlattice-like models containing at least two coupled interfaces [32,73,74]. In such algorithms, there are charge transfers and strong dipole-dipole interactions between the interfaces, inducing large artificial errors. In addition, on asymmetric interfaces, only relative stabilities [32] or average interface energies [73] could be obtained in such cases. Instead, our approach is able to isolate interfaces from each other, and the absolute interface energy of a single interface could be achieved.

Significant surface reconstructions with and without H will be calculated and discussed in this paper according to previous works and our calculated phase diagram shown later. The choices of low energy reconstructions without H follow previous investigations [53,57,69,70]. Specifically, on ZnO polar surfaces, although several different surface reconstruction models are proposed by both theoretical calculations and experiments [53,70], the surface energy differences among them are in the order of a few meV/Å$^2$ [70], suggesting that the (2×2) adatom reconstruction is adequate for our discussions here. For GaN, (2×2) adatom reconstructions were taken into consideration under the condition of N-rich [57]. Such reconstructions were suggested to be important under high temperature growth.[52]

As for the reconstructions with H, numerous possible H covered reconstructions should be considered. However, the lowest energy configurations must be those surfaces that satisfy ECR [52,56]. It demonstrated that the vibrational entropy of the surface plays a decisive role in the competition among various phases to achieve thermodynamic stability [53]. And when the temperature (and pressure) effects of the gas phase reservoirs, which arises from vibrational contributions to the entropy [56], are significant in high temperature growth, it would constrain the reconstructions within a limited number of configurations with less hydrogen atoms [56]. As a consequence, all the configurations of reconstructions will be shown later. What's more, except for those we calculated and discussed later, it showed in ref. [55] (see Fig. 13) that the surface reconstructions of (5×5) H4 and (5×5) H5 are also competing phases for the -c plane of ZnO. We will show later these phases would not affect our conclusion significantly.

Ideally, a systematic comparison among all possible surface configurations and a detailed surface phase diagram are sometimes used to demonstrate the problems on surfaces[51,56,75,76]. However, this is not easily defined when it comes to a heterogenous case, where more complex reconstructions than that in pure substance system may exist during the formation of heterostructures, especially for epitaxial films. Here we simplified the searching process and used the most energetic favorable configurations in ZnO or GaN to model the top surface of the heterostructures. We believed that our discussions on the selected detrimental reconstructions may still show the thermodynamic driving force to demonstrate the problem. In the unlikely case, even if we missed the lowest energy configuration in GaN, reconstructions with lower surface energies will not change our conclusion and are better for the realization of our strategies, as the inequality (1) showed.

The total energy calculations of bulks, slabs and clusters [36] were based on Density

Functional Theory [77,78] as implemented in VASP code [79], with a plane wave basis set [80,81] and PBE Generalized Gradient Approximation (GGA) as the exchange-correlation functional [82]. GGA functional severely underestimates the band gap of GaN and ZnO, which may lead to wrong energy eigen values or electron occupations in the surface states [57]. Therefore, we performed calculations using hybrid functional of Heyd, Scuseria, and Ernzerhof (HSE) [83,84] on slabs of polar surfaces and interfaces, as well as pseudo-molecules [36]. Slabs for surface reconstructions were performed on 2×2 slabs containing 10 bilayers, with 9×9×1 Gamma-centered k-point mesh for GGA and 3×3×1 for HSE. For the H involved reconstructed (0001) surface of ZnO, 3×3 slabs and 6×6×1 k-point mesh was applied for GGA. In the case of interface energy calculations, we construct the ZnO/GaN interface with 6 bilayers ZnO and 6 bilayers GaN on 1×1 slabs for ideal cases and 2×2 slabs for compensated cases. The k-point sampling is up to 15×15×1 and 7×7×1 for GGA, 7×7×1 and 3×3×1 for HSE. The slabs were separated by a vacuum of at least 15Å. All the atoms in the slab were allowed to relax until forces converged to less than 0.005eV/Å for GGA calculations and 0.01 eV/Å for HSE calculations. The energy cutoff of the plane-wave basis set was set to 500 eV for GGA and 400eV for HSE. We have done careful convergence tests for all of the settings aforementioned.

Calculations for the formation energies of related compounds are necessary for obtaining the thermodynamic phase diagram of interfaces shown later. However, there exists an intrinsic difficulty to correctly estimate the formation energy of the possible secondary phase of $Zn_3N_2$ near the interfaces, which is positive, for both our calculations and previous works, due to the intrinsic problems of exchange correlation functional [85]. And the calculated formation energies of compounds, such as Ⅱ-Ⅵ or Ⅲ-Ⅴ, are always significantly different from experimental ones [86], which are important to our further discussions. Therefore, we applied and strictly followed the fitted elemental-phase reference energies (FERE) method [86-88] to achieve relatively accurate formation energies. In the FERE method, for GGA, totally 12 elements and 44 compounds were included; for HSE, totally 9 elements and 24 compounds were included. Readers can refer to ref. [86-88] for more calculation details.

## III. Results and Discussions

First, we properly estimated the effect of secondary phases near the interface. These phases may significantly increase the interface energy and ruin the layer by layer growth mode. Therefore, to obtain sharp interfaces, we have to suppress the formation of any interlayers of possible secondary phases, including $Ga_2O_3$, $Ga_2ZnO_4$ and $Zn_3N_2$. The calculated formation energies of the important compounds by FERE method in this study are listed in **Table I**. The calculated results are well consistent with experimental values.

**TABLE I**

Calculated formation energies by FERE method, with both GGA and HSE, compared with experimental values.

|        | GGA (FERE) /eV | HSE (FERE) /eV | Exp. /eV |
|--------|----------------|----------------|----------|
| ZnO    | -3.53          | -3.73          | -3.62[89] |
| GaN    | -1.47          | -1.41          | -1.62[90] |
| Zn$_3$N$_2$ | -0.22     | -0.26          | -0.25[89] |
| Ga$_2$O$_3$ | -11.43    | -11.74         | -11.30[89] |

We obtain the corresponding phase diagram of the Zn, O, Ga, N compounds aforementioned, as shown in **Fig. 2**. The shadow part of the diagram is the chemical potential region, where only ZnO and GaN phases can form. Therefore, nearly O-poor and N-rich growth conditions are required to grow secondary-phase-free ZnO/GaN interfaces. Qualitatively, this phase diagram is nearly functional-independent. The growth window by HSE is slightly larger than that of GGA. To make a clear demonstration, in the following discussions, we directly compare the results along the four sampling points (e.g. A, B, C, D for GGA and A`, B`, C`, D` for HSE) in the two figures simultaneously. Additionally, with the existence of H atoms, according to the calculated results, H$_2$O formations is excluded because the chemical potential window of O-rich condition is needed for that.

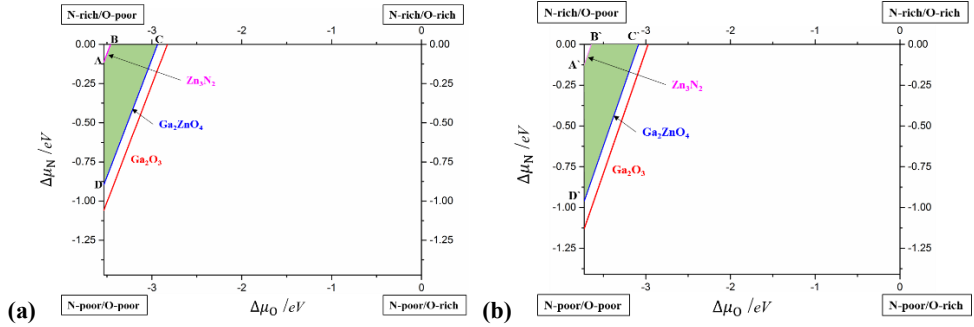

**FIG. 2**. Phase diagram of ZnO and GaN interfaces with **GGA** (a) and **HSE** (b) functionals. The shadow part is the stable chemical potential region where no secondary phases can form. In both of the figures, four sampling points are selected, denoted by A, B, C, D and A`, B`, C`, D`.

Next, we obtained the absolute surface energies of cleaved and reconstructed polar surfaces of ZnO and GaN. The surface energies of c and –c planes were calculated, as shown in **Fig. 3**. Generally, the energies calculated by HSE are larger than that by GGA. The overall shapes of both sets of data along the sampling points for GGA and HSE are quite similar. It indicates that there exist approximately constant shifts between the results based on the different functionals. And GGA largely underestimated the energies of cleaved surfaces. The errors of the GGA results are likely due to the fact that GGA functionals underestimate the energies with empty orbitals. For ZnO, the surface energies of reconstructed c and –c planes are close, -c plane slightly higher than c plane. The energy difference obtained from HSE is slightly larger than that obtained from GGA. And at A`, B` and D`, namely O extremely poor points, cleaved (0001) surface is slightly more stable than the (2×2) O adatom surface. In GaN results, the surface energy of reconstructed c plane is slightly higher. We found that this is different from the results shown on Fig.2 in ref. [57]. The major reason is due to the weak algorithm adopted in ref.[57]. The wedge structure in that paper is stressed, the steric effect of H is significant, and the self-consistence is poor [56]. Our approach yields an improved accuracy with the error about a few meV/A$^2$. Please check our early paper for detailed discussions[36,91].

In addition, we consider the configurations of sharp interfaces, where no interlayers can be formed in the growth window, as shown in **Fig. 2**. Under these conditions, we can then compare the absolute interface energies of different interface configurations. We estimated the interfaces energy of ZnO/GaN heterostructures using the lattice constant of ZnO and that of GaN. We found that the difference of interface energies between these two cases is less than 3 meV/Å$^2$. Therefore, such choices would not qualitatively affect our conclusions.

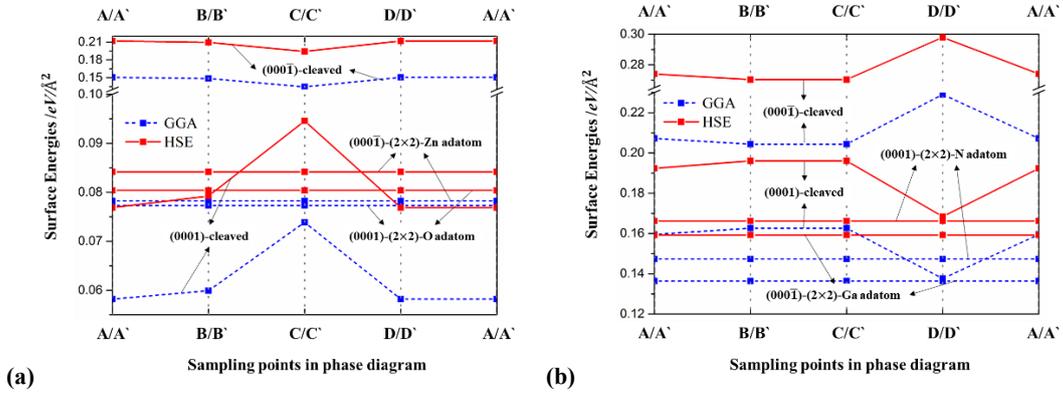

**FIG. 3**. Calculated absolute surface energies of cleaved and (2×2) reconstructed polar surfaces for ZnO (a) and GaN (b), along the sampling points in **Fig. 2**, with both GGA and HSE functionals.

Next, we considered different interfacial configurations. Experimentally, it is observed that the epitaxial crystal has the same polarity as the substrate without interface modifications and substrate pre-treatment [26,28,92]. Therefore, we first tried all the interface configurations without adlayers. Next, we also calculated adlayers inserted interfaces. There are totally 16 possible configurations. We found that the interfaces of 'N-Ga-O-Zn' and 'Ga-N-Zn-O', which preserve AB packing of wurtzite structure, are the most stable cases [93], as shown in **Fig. 4 (a)** and **(b)**. It indicates that when the epi-layers have the same polarity as the substrates do, the interfaces are relatively stable. Both interfaces were also speculated by experimental groups [92]. However, they may not be energetically favorable in all cases, because these interfaces can be negatively or positively charged. To further decrease the interface energy, it's possible to reconstruct the interface by charge compensations, as suggested in ref.[32]. We plotted our calculated structures, as shown in **Fig. 4 (c)** and **(d)**. Here we only consider the compensation of one layer in vicinity to the interfaces. If there are more stable interfacial configurations, our conclusion on the surfactant strategy shouldn't change because the inequality (1) still preserves.

The calculated absolute interface energies are shown in **Fig. 5**. For the cation-compensated interfaces, the absolute interface energies are low and almost functional-independent. While for the ideal interfaces, it seems that HSE results have a constant shift upward when they are compared with GGA results. At the points D/D`, GGA wrongly estimates that the ideal interface [N-Ga-O-Zn] energy to be negative, lower than that of the cation-compensated one. While HSE shows that the ideal interface energy is still higher than the compensated case. Therefore, in the following discussions, we only included the absolute interface energies of compensated cases.

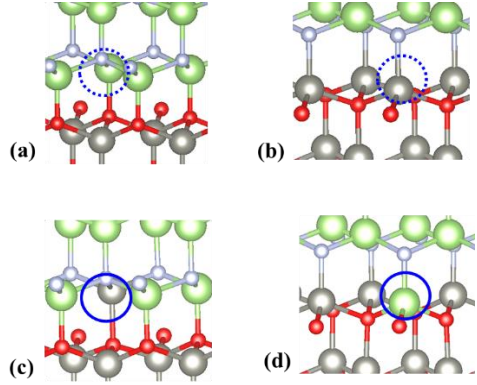

**FIG. 4**. Calculated interface configurations for ZnO/GaN heterostructures: (a) ideal [N-Ga-O-Zn], (b) ideal [Ga-N-Zn-O], (c) cation-compensated [N-Ga-O-Zn], and (d) cation-compensated [Ga-N-Zn-O]. (Green atoms for Ga, silver for N, grey for Zn and red for O) The circled positions are the sites of the replacement.

To determine the wetting conditions, we plugged in the surface energy of substrates, the interface energy and the surface energy of epi-layers to Eq. (1). We can conclude that:

$$\sigma_s[\text{GaN}(0001)_{\text{reconstructed}}] - \sigma_i[\text{N} - \text{Ga} - \text{O} - \text{Zn}] - \sigma_f[\text{ZnO}(0001)_{\text{reconstructed}}] > 0 \quad (3)$$

, as well as,

$$\sigma_s[\text{GaN}(000\bar{1})_{\text{reconstructed}}] - \sigma_i[\text{Ga} - \text{N} - \text{Zn} - \text{O}] - \sigma_f[\text{ZnO}(000\bar{1})_{\text{reconstructed}}] > 0 \quad (4)$$

. These results suggest that the ZnO thin films wet both c and –c directions of GaN substrates, but not vice versa. Also, it is very challenging to grow high quality GaN thin films on ZnO substrates because of a large difference of more than 80 meV/Å$^2$ between the surface energy of ZnO and that of GaN along c or –c directions. This is consistent with experimental observations[12,23-35]. Also, we found that this conclusion is functional-independent.

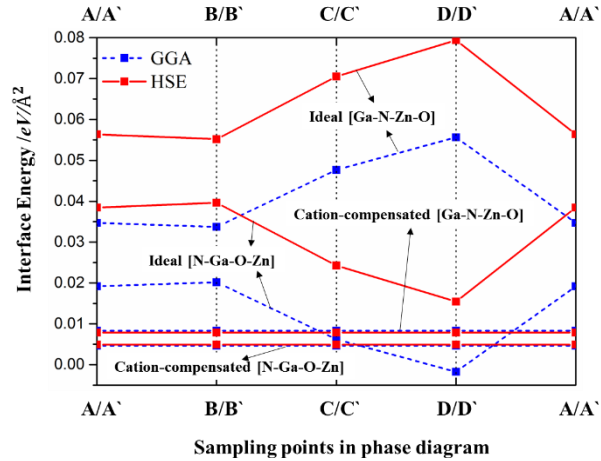

**FIG. 5**. Calculated absolute interface energies with the configurations shown in **Fig. 4**, along the sampling points in **Fig. 2**, by GGA and HSE functionals.

However, the intrinsic wetting conditions can't meet the industrial needs of cheap and lattice matched substrates for GaN. Therefore, it is desirable to overturn the wetting conditions.

One of the possible strategies would be surfactant assisted growth. Proper surfactants may tune the surface energies of ZnO and GaN to fulfill the wetting conditions, so that high quality GaN can be grown on ZnO substrate achieving the layer by layer growth mode. Here we found that H may be one of such surfactants. Hydrogen may attach to the polar surfaces and form H-involved reconstructions of ZnO and GaN to lower the overall surface energies. According to our calculated growth window, we have to grow the heterostructure under N rich, O poor conditions. Therefore, following early experimental and theoretical results, we adopted the configurations of H adsorbed surfaces with lowest energies, as shown in **Fig. 6**. In this discussion, these reconstructions should also be compared with the adatom reconstructions. All of the configurations except (b) and (c) in **Fig. 6** satisfy ECR.

For the c plane, the calculated results show that the surface energy of GaN is always larger than that of ZnO, the difference still larger than 30 meV/Å$^2$. In this way, it is impossible to change the wetting condition. So, we discarded this case.

For the reconstructions of -c planes, different numbers of H atoms are involved on the most stable surface configurations, 2 H for ZnO and 3 H for GaN, shown in **Fig. 6**. Initially, we kept constant H concentration. Therefore, the relative differences of the surface energies for ZnO and GaN are H chemical potential independent. We considered two extreme cases and examined them through calculations by GGA functional. First, when both surfaces adsorbed 2 H atoms, the absolute surface energies of ZnO is much lower than that of GaN, with a difference more than 23 meV/Å$^2$. In addition, when both surfaces adsorbed 3 H atoms, the absolute surface energies of GaN is much lower than that of ZnO, with the difference more than 27

meV/Å$^2$. In this case, it is likely for GaN to wet ZnO on the -c plane. Therefore, it means that it is necessary to precisely control H concentration to tune their wetting conditions. However, in realistic growth conditions, especially under thermodynamic equilibrium near the surface or interface, it may be difficult to maintain a constant adsorbed-H concentration on different material surfaces, because of different global minimum configurations. And thus, the key to this problem is whether we are able to reduce the surface energies of GaN -c plane to make it even lower than that of the most stable configuration of ZnO -c plane, with the same H chemical potential condition. So, the safest way to demonstrate our method is to compare their most stable configurations, i.e. **Fig. 6(d)** and **Fig. 6(e)**. Thus, the surface energy difference is H chemical potential dependent, due to different adsorbed H atom numbers. Therefore, accurately controlling H chemical potential through $T$ and $p$ is especially crucial.

Then, we focused on the –c planes of ZnO and GaN, shown in **Fig. 6(d)** and **Fig. 6(e)**, which are configurations fully satisfying ECR. The following relationship is expected:

$$\sigma_s[\text{ZnO}(000\bar{1})_{\text{reconstructed}}] - \sigma_i[\text{N} - \text{Ga} - \text{O} - \text{Zn}] - \sigma_f[\text{GaN}(000\bar{1})_{\text{reconstructed}}] > 0. \quad (5)$$

Here the surface energies of H-involved surfaces are largely dependent on the H free energy, namely:

$$\sigma_{s-H} = \frac{1}{\alpha_{\text{surface}}}(E_{\text{slab}} - \sum_i n_i \cdot \mu_i - n_{H_X}\hat{\mu}_{H_X} - n_H\mu_H), \quad (6)$$

where $E_{\text{slab}}$ is the total energy of the slab model, $\alpha_{\text{surface}}$ is the surface area of the reconstructed surface, $n_{H_X}$ is the number of corresponding pseudo-H atoms used to passivate the bottom, and $\hat{\mu}_{H_X}$ is the PCPs of corresponding pseudo-H. More importantly, $n_H$ is the number of H atoms involved in the reconstructions, and $\mu_H$ is the chemical potential of H atoms, which can be expressed as:

$$\mu_H = \frac{1}{2}(E_{H_2} + 2\Delta\mu_H), \quad (7)$$

where $E_{H_2}$ is the total energy of H$_2$ at zero temperature (derived from DFT calculations), $\Delta\mu_H$ is the term of the chemical potentials relative to the total energy of the isolated molecule, which is determined by the gas atmospheric temperature and pressure. Quantitatively, this formula can be solved theoretically by the partition functions of diatomic idea gases [56,62,94], which includes a sum over translational, rotational, and vibrational states [95]. As suggested by early literature, the vibrational state of H$_2$ is the main contribution under high growth temperature [56,62]. Besides this method, we noticed that another quantifiable relation of the quantity $\Delta\mu_H$ is[95]

$$2N\Delta\mu_H = G, \quad (8)$$

where $N$ is the number of H$_2$ molecules, G is the Gibbs free energy of H$_2$ (gas) in reference to zero temperature. Therefore, in this paper, we refer to the experimental data of H$_2$(gas) Gibbs free energy G [96], including both enthalpy and entropy contributions, to obtain the quantitative relation of the expression aforementioned.

Therefore, when the Eq. (5) is satisfied, we can obtain $\Delta\mu_H$ through substituting Eq. (6) and Eq. (7) into Eq. (5), shown in **Table II**. Also, the corresponding absolute surface energies of H involved surfaces are calculated according to Eq. (6), shown in **Table II**. For the GaN H-involved reconstructed $(000\bar{1})$ surface, the surface energies are all lower than the (2×2) adatom one, at the four sampling points, both GGA and HSE. While for ZnO, only at C/C` and D/D` points, the H-involved reconstructed surface is favorable than the (2×2) adatom one. Generally, GGA and HSE give quite consistent predictions: at all of the four sampling points H-adsorbed GaN -c surface has lower surface energies. It means that in the region near these sampling points, hydrogen surfactants would turn over the intrinsic wetting condition, making inequality (5) possible, when the growth temperature and substances' partial pressures are satisfied. Additionally, for the possible surface reconstructions of (5×5) H4 and (5×5) H5 on the -c plane of ZnO, at A/A` and B/B`, when $\Delta\mu_H$ continuously increased by ~200 meV, the reconstruction of (d) in **Fig. 6** would took over the surface. So, these (5×5) phases actually dominate a quite small region in the whole surface phase diagram. Also, such large period reconstructions could hardly exist in high temperature epitaxial growth processes. It means that these (5 × 5) reconstructions may not qualitatively and quantitatively affect our conclusion in a large degree. For the same reason, it is also reasonable to ignored such reconstructions at C/C` and D/D`. As a result, under the calculated range of H chemical potentials, all of the reconstructions (**Fig. 6 (d), (e)** and (2×2) adatom) discussed above are sufficient to demonstrate the whole picture.

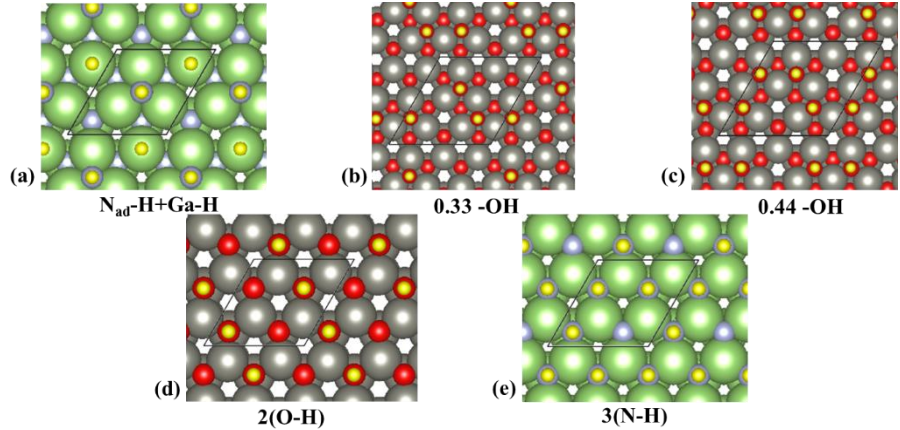

FIG. 6. Hydrogen involved (0001) and $(000\bar{1})$ surface reconstructions for GaN (a, e) and ZnO (b, c, d). The yellow atoms are H. Other atoms follow the notations in **Fig. 4**. The parallelogram in each configuration shows the smallest period of the surface structure.

**TABLE II**

Hydrogen chemical potential relative to the total energy of the isolated molecule, denoted as $\Delta\mu_H$, when wetting conditions are satisfied critically. It shows along the sampling points illustrated in **Fig. 2**. The surface energies with the configurations shown in **Fig. 6 (d)** and **(e)** at that stage are also calculated, denoted as $\sigma_{ZnO-H}$ and $\sigma_{GaN-H}$. And $\sigma_{ZnO}$ represents the absolute surface energy of the (2×2) reconstructed surface of ZnO. Here only lowest surface energies of ZnO and GaN are shown in this table. Predicted maximum growth temperatures (under the pressure of 1 bar) for possible sampling points are denoted as *T*. The data with underlines are those that we should

|  |  | A | B | C | D |
|---|---|---|---|---|---|
| **GGA** | $\Delta\mu_H$ /eV | -1.065 | -1.099 | -1.070 | -0.782 |
|  | $\sigma_{ZnO-H}$ /meV/Å$^2$ | - | - | **76.3** | **76.8** |
|  | $\sigma_{GaN-H}$ /meV/Å$^2$ | **73.6** | **73.6** | **71.3** | **71.9** |
|  | $\sigma_{ZnO}$ /meV/Å$^2$ | **78.5** | **78.5** | - | - |
|  | T /K ($p$ ~1 bar) | ~1400 | ~1400 | ~1400 | ~1100 |
|  |  | A` | B` | C` | D` |
| **HSE** | $\Delta\mu_H$ /eV | -1.106 | -1.149 | -0.995 | -0.675 |
|  | $\sigma_{ZnO-H}$ /meV/Å$^2$ | - | - | 71.4 | 71.4 |
|  | $\sigma_{GaN-H}$ /meV/Å$^2$ | 79.2 | 79.2 | 66.4 | 66.4 |
|  | $\sigma_{ZnO}$ /meV/Å$^2$ | 84.2 | 84.2 | - | - |
|  | T /K ($p$ ~1 bar) | ~1400 | ~1400 | ~1300 | ~1000 |

In this way, we then referred to the corresponding experimental Gibbs free energies data G of H$_2$ expressed as a function of temperature and pressure [96]. Therefore, when $\Delta\mu_H$ equals to the corresponding $G$, according to Eq. (8), we can relate it to a given temperature $T$ and pressure $p$. In principle, for a certain $\Delta\mu_H$, a lower pressure required a lower temperature. Here, the pressure was set as 1 bar, which is almost the pressure for typical OMVPE growth. While for the typical H$_2$ pressure of MBE growth ~10$^{-12}$ atm, the critical temperature would be far lower than the typical value of 1000K [66], so we neglect related discussions. In real experiments, the H$_2$ pressure may be higher because of the unavoidable passivation of bulk or interface defects. The extra incorporation of H can be thermally annealed later [61] . Therefore, our results give a lower limit of H$_2$ pressure. The critical temperatures, i.e. the maximum growth temperature by using H surfactant effect for the –c plane of GaN wetting that of ZnO, are estimated and shown in **Table II**. For all of the sampling points, the differences of $\Delta\mu_H$ between GGA and HSE are quite small, within 110 meV. So, the estimated critical temperatures of them are almost the same, the differences within 100 K. For A/A` B/B` and C/C`, the critical temperatures are about 1300 ~ 1400 K. It should be noted that for D/D`, the critical temperature is around 1000 ~ 1100 K, so the condition of D/D` cannot be applied directly with standard OMVPE temperature T = 1300 K.

In addition, Ga-rich condition is mainly applied in most growth techniques to yield good crystal quality and flat surfaces. It means that the growth condition near D/D` is preferred from the crystal quality point of view. Also in such region near Ga-rich condition, those detrimental deep defects such as N$_{Ga}$, V$_{Ga}$ or V$_{Ga}$-H related complexes can be suppressed to a certain degree. The interplay of the H passivation between surfaces and point defects nearby is also an interesting problem to be discovered, yet out of scope here. Additionally, the growth temperature strongly influences the surface morphology and affects the optical properties of the GaN epitaxial films[97]. For other three points, especially A/A` and B/B`, i.e. the growth window near N-rich and O-poor, wider flexibility of controlling the growth temperature can be achieved. The relatively lower crucial temperature at point D/D` is also beneficial to the stability of pre-grown ZnO buffer layers or substrates.

Moreover, In the recent years, the topic of growing N-polar GaN with comparable high crystal quality to Ga-polar GaN has attracted much attention[98-100], due to the special advantages of N-polar thin films over Ga-polar ones[43,101-104]. Mg, In and Si are used to assist to get rid of the disadvantages of the N-polar film growth on traditional sapphire and SiC substrates[99,105]. In our predictions, N-polar GaN of high crystal quality is very likely to be grown on ZnO O-polar substrate. Therefore, our results may provide a new method to achieve this goal, and further investigations on the effect of growth conditions based on our results are necessary. Also, in addition to H, other group I or valence I elements can be attractive surfactant candidates that can be investigated in the future.

## IV. Conclusion

In summary, hydrogen surfactant effect was investigated by determining the wetting conditions of ZnO/GaN heterostructures at the first time, by GGA and HSE functionals. We also proposed a new strategy to accurately estimate the interface energy. Then the phase diagram of sharp interface is obtained. Within this nearly N-rich and O-poor growth window, the absolute surface energies of the related reconstructed surfaces are obtained, as well as the absolute interface energies with possible configurations. It shows that intrinsically it would be challenging to grow high quality GaN thin films on ZnO substrates. To turn over the wetting condition, we proposed hydrogen as surfactants during the growth due to its advantages over other elements. When H-involved surface passivation is considered, with detailed analysis of atomistic thermodynamics, we predicted and verified that H may be a good surfactant tuning the epitaxial growth mode and making it possible for GaN to achieve layer by layer growth mode on the $(000\bar{1})$ surface of ZnO. The proper growth conditions are also predicted. Our work may benefit the applications of GaN and related ZnO/GaN heterostructure devices. Making ZnO as possible substrates or interlayer during GaN growth may reduce the usage of expensive substrates or gallium elements. Also, such surfactant effect analysis shed light on further investigations on the surfactant-assisted heterostructures growth especially for complex compounds.

## Acknowledgements

Part of the computing resources was provided by the High-Performance Cluster Computing Centre, Hong Kong Baptist University. This work was supported by the start-up funding, HKRGC funding with the Project code of 2130490, and direct grant with the Project code of 4053233, 4053134 and 3132748 at CUHK.